\documentclass[a4paper]{article}

\usepackage{icrc2013}

\title{On the redshift of the gamma-ray blazar PKS 0447-439: Optical spectroscopy using Gemini observations with high S/N ratio}

\shorttitle{On the redshift of PKS 0447-439: Spectroscopy using Gemini observations with high S/N ratio}

\authors{
A.C. Rovero$^{1}$,
C. Donzelli$^{2,3}$,
H. Muriel$^{2,3}$,
A. Cillis$^{1}$
A. Pichel$^{1}$
}

\afiliations{
$^1$ Instituto de Astronom\'ia y F\'isica del Espacio (CONICET-UBA), Buenos Aires, Argentina \\
$^2$ Instituto de Investigaciones en Astronom\'ia Te\'orica y Experimental (CONICET), C\'ordoba, Argentina \\
$^3$ Observatorio Astron\'omico de C\'ordoba, C\'ordoba, Argentina \\
}

\email{rovero@iafe.uba.ar; 
charly@oac.uncor.edu}

\abstract{
The BL-Lac blazar PKS 0447-439 was detected at very high energy gamma-rays by HESS following the discovery by Fermi-LAT. The lack of both emission and absorption lines in BL-Lacs make the estimation of their redshifts very difficult.  
Modeling the drop in gamma-ray spectra it was possible to have an estimation of redshift for PKS 0447-439 of z$\sim$0.2, which is
compatible with the value z=0.205 reported by the identification of Ca II absorption lines in optical spectra.
By the identification of a weak line of Mg II using spectra with average signal-to-noise S/N$\sim$80, it has been recently reported a lower limit for the redshift of this blazar of z$\leq$1.246.
Triggered by this controversy, we have proposed new optical observations with the Gemini South telescope to perform further spectroscopic analysis with very high S/N ratio ($\sim$200-500).
In this work we present a new optical spectrum of PKS 0447-439, and report on the analysis and results of such observations. 
Even with this significantly high quality signal we were not able to identify any spectral features to allow an estimation of the redshift. In agreement with other recent studies, we identify the Mg II line reported previously as originated in the Earth's atmosphere, and conclude the lower limit of the redshift is incorrect. More interestingly, we could not identify the Ca II absorption lines used to report a redshift of 0.205.
}


\begin{document}
\maketitle

\section{Introduction}

Gamma-ray radiation in the range of very-high-energy (VHE; E$>$100 GeV) is strongly attenuated by the photon-photon interaction with the extragalactic background light (EBL) and the cosmic microwave background (CMB). As a consequence, all discovered VHE sources are relatively close (z$<$0.6). The most popular extragalactic objects in VHE catalogs are blazars, mostly BL-Lacertae (BL-Lac). The lack of both emission and absorption lines in BL-Lacs make the estimation of their redshifts very difficult. An alternative method is to model the drop in the spectral index from high-energy (HE; E$>$100 MeV) to VHE due to the photon-photon interaction, for which both the spectrum at HE and VHE have to be known and a distribution of the low energy photon field has to be assumed.

The BL-Lac blazar PKS 0447-439 is one of the brightest HE sources first detected by Fermi-LAT \cite{bib:abdo}. It was also detected at VHE by H.E.S.S. with a very soft derived spectrum (photon index 4.3) and no indication of break or curvature \cite{bib:zech}. Modeling the drop in the spectral index from HE to VHE, and assuming an EBL density \cite{bib:franceschini}, it was possible to derive an estimation of redshift for PKS 0447-439 to be z$\sim$0.2 \cite{bib:prandini}, which is compatible with the upper limit z$<$0.53 obtained by a similar procedure \cite{bib:zech}. 
These results are in agreement with the spectroscopic value z=0.205 reported in 1998 using data taken with the CTIO 4 m telescope, in which the identification of weak Ca II absorption lines was claimed \cite{bib:perlman}. A lower limit of z$>$0.176 for the redshift of PKS 0447-439 was also estimated from the optical V magnitude \cite{bib:landt2}.

Recently, a significantly higher lower limit of z$>$1.246 for the redshift of PKS 0447-439 was announced \cite{bib:landt}. This spectroscopic redshift is based in the identification of the line Mg II $\lambda$2800 doublet in absorption using two spectra from the CTIO 4 m and ESO-NTT 3.6 m telescopes, with average S/N$\sim$80. A high redshift like this for a VHE source would imply either that the relevant absorption processes of gamma-rays are not well understood or that the EBL is dramatically different from what is believed today. An alternative non exotic explanation for this high redshift was proposed \cite{bib:aharonian}, arguing that the distant TeV blazar emission could be compatible with secondary photons produced by energetic protons from the blazar jet propagating over cosmological distances almost rectilinear.

While this controversy was rising, a spectrum taken in December 2011 with the X-Shooter on the VLT was presented in July, 2012 \cite{bib:pita}. Also, a new MagE/Magellan spectrum of PKS 0447-439 with S/N$\sim$50-150 from observa-tions taken on July 12, 2012, was presented \cite{bib:fumagalli}. They both confirmed the presence of the absorption line at 6280 \AA, which could result in a high redshift value if identified as Mg II $\lambda$2796.82. However, they associated this line with a known telluric absorption line, invalidating the claim that this is a very distant blazar. No other spectral features were detected so, in neither of these publications a redshift for PKS 0447-439 was established.

At the same time, we submitted a proposal in April, 2012, for the observation of this blazar with the Gemini South telescope in Chile. In this work we report the analysis and the results of such observations taken on November 21, 2012. We present a new spectrum of PKS 0447-439 from the data acquired with the GMOS-S spectrometer with a significantly higher signal-to-noise ratio, as well as the results of the search for lines previously reported for this blazar.

\begin{figure*}[!ht]
  \centering
  \includegraphics[width=0.8\textwidth]{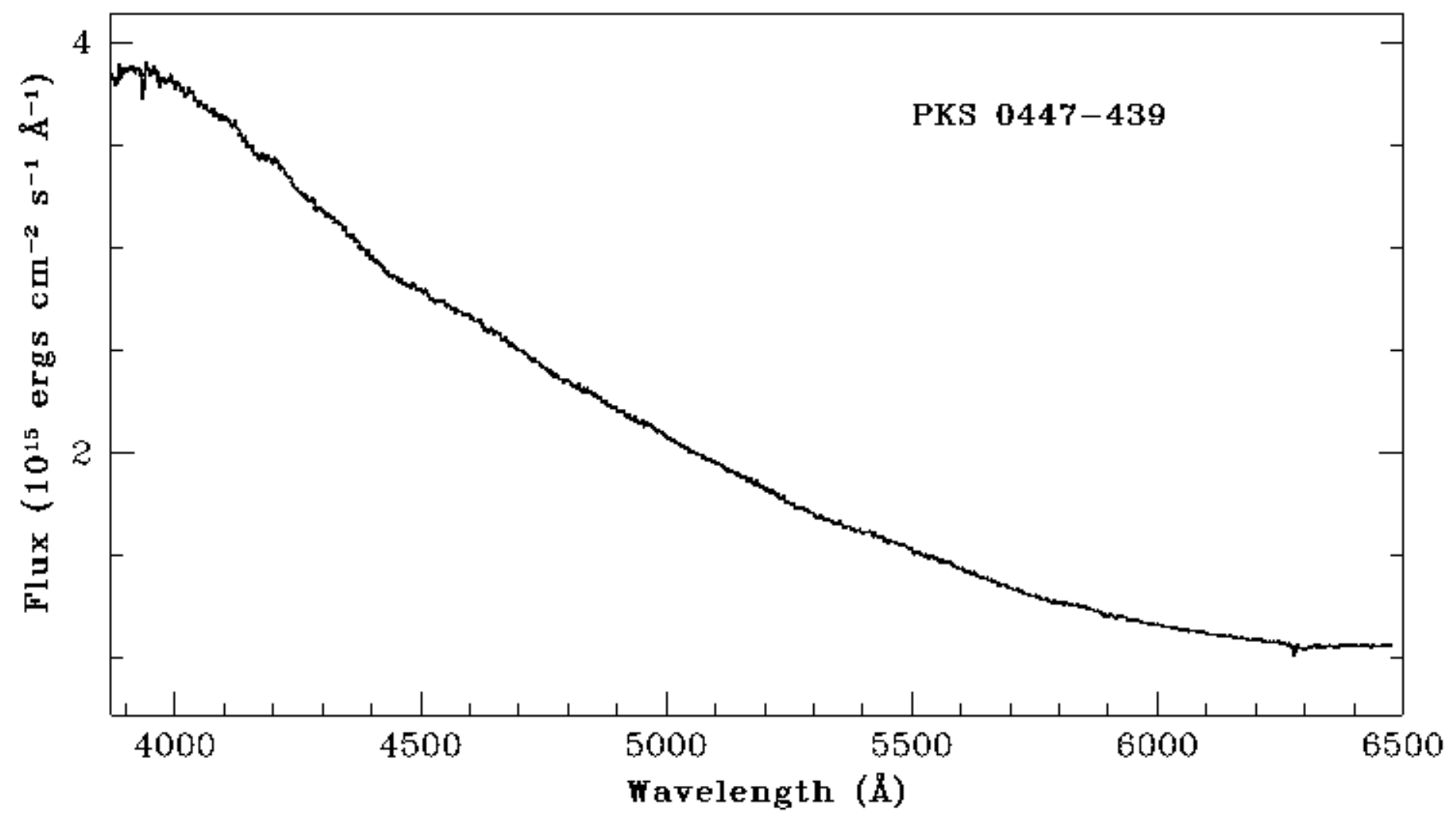}
  \caption{Observed spectrum of PKS 0447-439}
  \label{fig:espectro}
\end{figure*}

\section{Observations and Data Reduction}

Spectra for PKS 0447-439 and twenty one other ob\-jects around it were obtained with the 
Gemini Multi-Object Spectrograph (GMOS), program GS-2012B-Q25 (PI A.C. Rovero). A multislit mask was created
for this purpose using a pre-image provided by Gemini. We placed one field centered on
PKS 0447-439, covering a region of $5\times5$ arcmin$^2$, and selected other extragalactic objects
in the field to characterize the PK 0447-439 environment and to search for other possible associations
with the gamma-ray source. These objects will be analyzed in a forthcoming paper. 

The spectroscopic data was acquired in queue mode on November 21, 2012, using the multislit 
mask. The grating in use was the B600$\pm$G5323 that has a ruling density of 
600 lines/mm. Three exposures of 900 s each through a 1.0$^{\prime\prime}$ slit were obtained with the central
wavelengths of 497 nm, 502 nm, and 507 nm. Science targets have thus a total exposure time of 0.75 hours.
Observations were taken at airmass 1.25 with a seeing of 0.9$^{\prime\prime}$.
Flatfields, spectra of the standard star $LTT\;7987$, and the copper-argon 
$CuAr$ lamp were also acquired to perform flux calibration. A binning of $2\times2$
was used, yielding a scale of 0.1456 arcseconds per pixel and a theoretical 
dispersion of $\sim0.9$ \AA~ per pixel. 

All science and calibration files were retrieved from the Gemini Science Archive 
hosted by the Canadian Astronomy Data Center. The data reduction described below
was carried out with the Gemini IRAF package. Flatfields were derived with the
task {\sc gsflat} and the flatfield exposures. Spectra were reduced using 
{\sc gsreduce}, which does a standard data reduction, i.e. it performs bias, 
overscan, and cosmic rays removal as well as the flatfielding 
derived with {\sc gsflat}. GMOS-South detectors are read with three 
amplifiers and generate files with three extensions. The task {\sc gmosaic} 
was used to generate data files with a single extension. The sky level was 
removed interactively using the task {\sc gskysub} and the spectra were 
extracted using {\sc gsextract}.

Flux calibration was performed using the spectra of the standard star $LTT\;7987$, 
acquired with an identical instrument configuration. Spectra of $CuAr$ lamps were 
obtained immediately after the observation of science targets and were used 
to achieve wavelength calibration using the task {\sc gswavelength}. The sensitivity
function of the instrument was derived using {\sc gsstandard} and the reference file
for $LTT\;7987$ was provided by the Gemini observatory.
Science spectra were flux calibrated with {\sc gscalibrate} which uses the 
sensitivity function derived by {\sc gsstandard}.

The redshift of targets in the field were calculated using the IRAF task {\sc fxcor} that
computes radial velocities by deriving the Fourier cross correlation between two spectra. 
As a reference spectrum we use data of the Galactic globular cluster BH 176 taken 
in the same GMOS configuration during a previous Gemini run \cite{bib:davoust}.\\

\section{Results}

Figure \ref{fig:espectro} shows the observed spectrum of PKS 0447-439 after data reduction and calibration.
The spectrum covers the range 3870-6475 \AA, in which we have determined a signal to noise ratio for the continuum
ranging from 200 at 4000 \AA∼ to 500 at 6000 \AA.\\


\subsection{spectral lines}

In the observed spectrum of PKS 0447-439 we clearly see the Galactic Na I absorption lines at 5891.6 \AA∼, 5894.1 \AA∼,
and 5897.6 \AA, and the Galactic Ca II H+K absorption lines at 3934.7 \AA∼ and 3969.6 \AA.  

\begin{figure}[!h]
  \centering
  \includegraphics[width=0.45\textwidth]{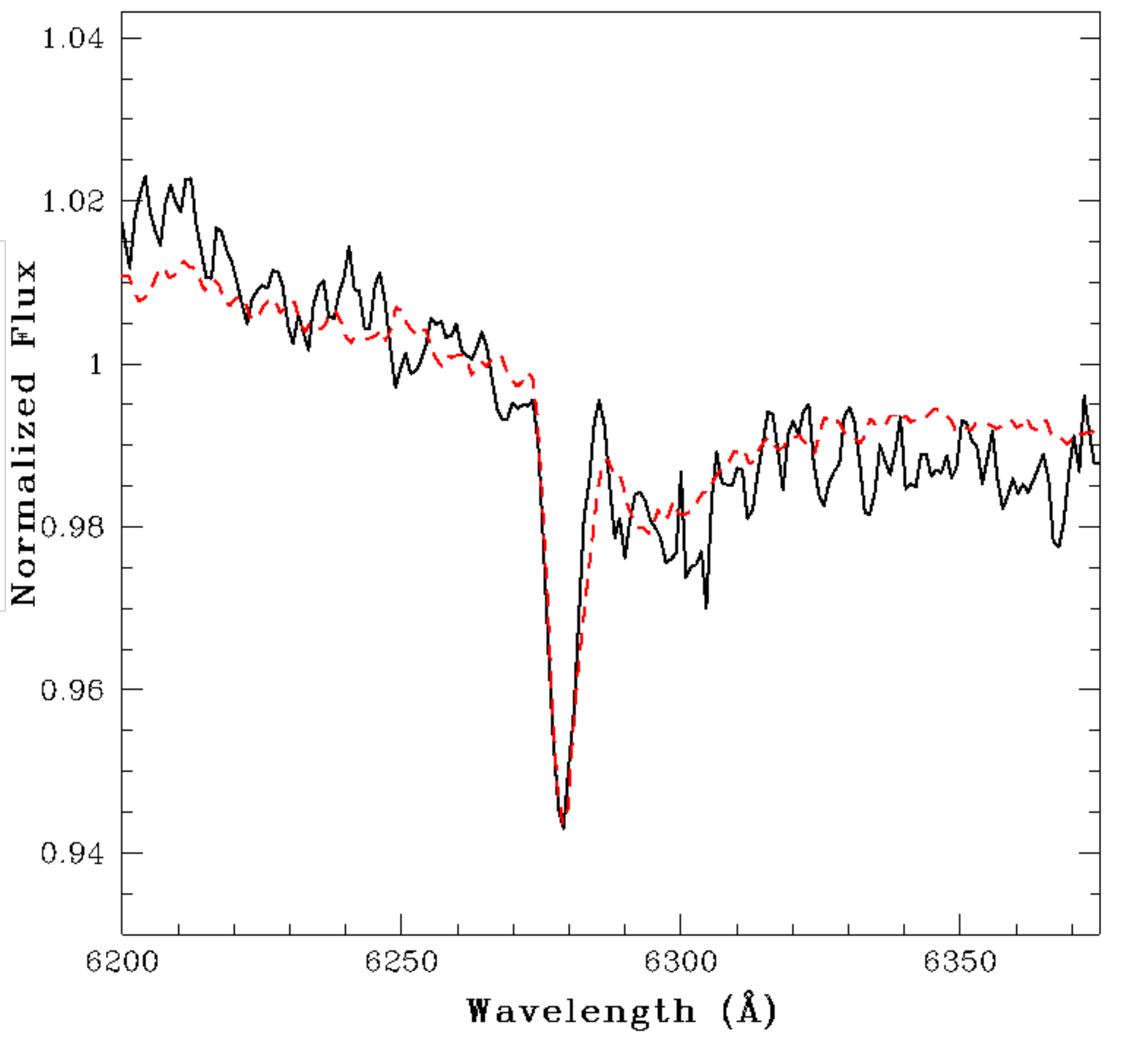}
  \caption{Absorption line at 6279 \AA∼ for both PKS 0447-439 (solid line) and the standard star (dashed line),
after flux normalization.}
  \label{fig:mg}
\end{figure}

We have also identified in the spectrum of PKS 0447-439 the absorption line at 6279 \AA, which led to the controversy
referred to above. This line is also present in the standard star spectrum and in some of the other spectra observed in our
program, which we interpret as originating in the Earth's atmosphere, in agreement with other recent conclusion \cite{bib:pita}  \cite{bib:fumagalli}.
Figure \ref{fig:mg} shows the perfect matching of the 6279 \AA∼ absorption line from both PKS 0447-439 and the standard star spectra, after flux normalization.

\begin{figure}[!h]
  \centering
  \includegraphics[width=0.45\textwidth]{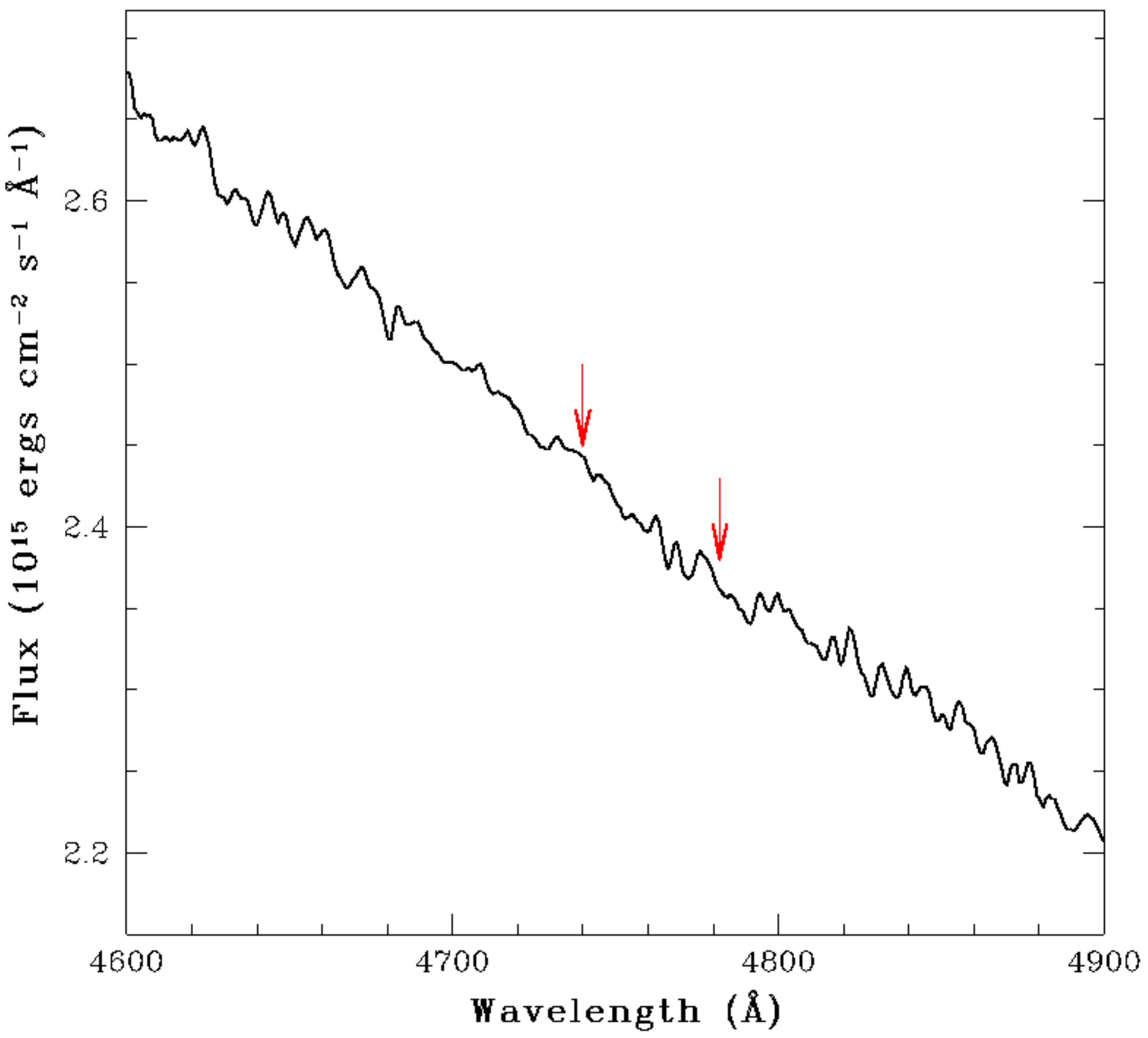}
  \caption{Region of the spectrum of PKS 0447-439 where the Ca II H+K absorption lines were identified at z=0.205 by
Perlman et al. \cite{bib:perlman}. The arrows indicate the position where the lines would be expected.}
  \label{fig:ca}
\end{figure}

Finally, we have searched the spectrum of PKS 0447-439 in the region where the Ca II H+K absorption lines were identified at
z=0.205 \cite{bib:perlman}. Figure \ref{fig:ca} shows the region from 4600 \AA∼ to 4900 \AA, with arrows at the position
where the lines would be expected for this redshift. We do not see these lines at a signal to noise ratio S/N$\sim$250.

\subsection{Redshift determination}

Within the range of wavelength covered by the new optical spectrum for PKS 0447-439 we have not identified any spectral
line from any chemical element other that those interpreted as either Galactic or terrestrial. Alternatively, we checked
visually possible evidences of the presence of ten absorption and six known emission lines with rest frame wavelengths
in the range 2800 \AA∼ to 6800 \AA. This was done in the redshift range z=0.0-1.5. No significant or marginal evidence
of coincidences between these commonly observed lines and the spectra of PKS 0447-439 was found. Consequently, no 
spectroscopic estimation for the redshift of this balzar could be determined in this work.

\section{Summary}

We observed the VHE blazar PKS 0447-439 with the Gemini South telescope on November 21, 2012, and performed
spectroscopic analysis with
very high signal to noise ratio (S/N 200 at 4000 \AA∼ to 500 at 6000 \AA).
Even with this high value of S/N we were not able to identify any spectral features to allow an estimation of the redshift.
Only Galactic and telluric spectral lines were identified. We also detected the absorption line at 6279 \AA∼ that was
identified by Landt \cite{bib:landt} as Mg II $\lambda$2800 doublet on which the report of a high redshift was based. However, we have
also observed this line in the spectrum of the standard star and concluded that this feature is originated in the Earth's
atmosphere. More interestingly, despite the high S/N of our observation we could not identify the Ca II absorption
lines used by Perlman et al. \cite{bib:perlman} to report a redshift of 0.205.\\

\vspace*{0.5cm}
\footnotesize{{\bf Acknowledgment:}{
This work is based on observations obtained at the Gemini Observatory, which is operated by the Association of Universities
for Research in Astronomy, Inc., under a cooperative agreement with the NSF on behalf of the Gemini partnership:
the National Science Foundation (United States), the National Research Council (Canada), CONICYT (Chile), the Australian 
Research Council (Australia), Minist\'{e}rio da Ci\^{e}ncia, Tecnologia e Inova\c{c}\~{a}o (Brazil) and Ministerio de
Ciencia, Tecnolog\'{i}a e Innovaci\'{o}n Productiva (Argentina). The following authors are members of ``Carrera del Investigador Cient\'ifico'' of CONICET: ACR, CD, HM and AC.}}

\end{document}